\documentclass{optica-article}

\journal{opticajournal} 

\articletype{Research Article}

\usepackage{amsmath}
\usepackage{amsfonts}
\usepackage{graphicx}
\usepackage{booktabs}
\usepackage{float}
\usepackage{xcolor}
 \usepackage{subcaption}
 \usepackage{hyperref}

\begin{document}
\title{Sub-picosecond inter-core skew characterization in multicore fibers via Hong--Ou--Mandel interference}

\author{L.~Lira Tacca,\authormark{1,2}
L.~Marques Fagundes,\authormark{3}
M. Morales Lillo,\authormark{1,2}
M. Navarro,\authormark{1,4,5}
I. Machuca,\authormark{1,2}
S. G\'omez,\authormark{1,6}
G.~H.~dos Santos,\authormark{1,2}
J.~Cari\~ne,\authormark{7}
G.~Saavedra,\authormark{8}
E.S.~G\'omez,\authormark{1,2}
G.~Lima,\authormark{1,2}
and S.~P.~Walborn\authormark{1,2,*}}

\address{%
  \authormark{1}Departamento de F\'isica, Universidad de Concepci\'on,
  160-C Concepci\'on, Chile\\
  \authormark{2}ANID -- Millennium Science Initiative Program --
  Millennium Institute for Research in Optics,
  Universidad de Concepci\'on, 160-C Concepci\'on, Chile\\
\authormark{3}Departamento de F\'isica, Universidade Federal de Santa Catarina,
  CEP 88040-900, Florian\'opolis, SC, Brazil\\
  \authormark{4}ICFO - Institut de Ciencies Fotoniques, The Barcelona Institute of Science and Technology, 08860 Castelldefels, Barcelona, Spain\\
\authormark{5}Luxquanta Technologies S.L., Av. Joan Carles I, 30,
1$^\circ$1$^\mathrm{a}$. 08908 L'Hospitalet de Llobregat,
Barcelona, Spain\\
\authormark{6}Dipartimento di Fisica, Sapienza Universit\`a di Roma,
Piazzale Aldo Moro 5, I-00185 Roma, Italy\\
  \authormark{7}Department of Electrical Engineering, Universidad Cat\'olica
  de la Sant\'isima Concepci\'on, Concepci\'on, Chile\\
  \authormark{8}Department of Electrical Engineering, Universidad de Concepci\'on,
  Concepci\'on, Chile}

\email{\authormark{*}swalborn@udec.cl}

\begin{abstract*}
Inter-core skew (ICS), the differential group delay between cores of a
multicore fiber (MCF), is a critical parameter for both classical
space-division multiplexed communications and quantum photonic networks.
We present a high-precision measurement of ICS in a commercially available
four-core fiber using two-photon Hong--Ou--Mandel (HOM)
interference in a fiber-integrated $4\times4$ multiport beam
splitter. By extracting the center position of HOM interference dips and
peaks across all twelve core-pair combinations, we obtain individual
ICS values with a demonstrated precision of
$\pm0.11\,$ps, limited by the delay-stage positioning uncertainty. The root-mean-square ICS grows as
$\sigma_\tau(L) = \kappa\sqrt{L}+c$ with
$\kappa = 48.7 \pm 2.5\,\mathrm{ps}/\!\sqrt{\mathrm{km}}$ and
$c = 9.76 \pm 1.2\,$ps, over fiber lengths from $7.7\,$m to $1300\,$m.
This first direct validation of the stochastic random-walk scaling across
a length range spanning laboratory to field-deployed scales was made
possible by HOM's immunity to first-order path fluctuations, which
renders classical interferometric methods impractical for long installed
fibers. The demonstrated $\pm0.11\,$ps precision represents a
$\sim\!180$-fold improvement over correlation optical time-domain
reflectometry (C-OTDR), the standard method for long-fiber ICS
characterization. Fisher information analysis establishes a fundamental
Cram\'er--Rao precision limit in the femtosecond range, indicating
further improvement is achievable with better delay control. These
results establish a practical platform for characterising timing
uniformity in MCF-based networks for both quantum and classical
space-division multiplexed applications.
\end{abstract*}


\section{Introduction}
\label{sec:intro}

Multicore fibers (MCFs) have emerged as a key enabling technology for
next-generation optical communication and quantum networking, offering
a scalable route to increased capacity through space-division
multiplexing (SDM) while distributing quantum states across spatially
distinct channels within a single cladding~\cite{richardson2013space,
saitoh2016multicore}. In the classical domain, MCFs support
petabit-class SDM transmission~\cite{Luis25}, joint-core coherent detection, and
radio-over-fiber applications demanding precise timing synchronization
across cores~\cite{richardson2013space,alfredsson2019performance,
cai2024impact}. MCF deployment has progressed to practical submarine systems, including a 4-core submarine-cable prototype with integrated multicore amplification~\cite{Takeshita23}, a deployed 7-core link carrying 410.5\,Tbit/s over 140\,km~\cite{Chen2026}, and recent work on key implementation building blocks~\cite{Nakanishi24}. In the quantum domain, the common cladding provides
intrinsic phase stability for high-dimensional entanglement
distribution~\cite{gomez2017,orttega2021}, parallel quantum
channels~\cite{canas2017,Dynes:16}, and fiber-integrated multiport
beam splitters~\cite{xavier2020quantum,carine2020multi}. Core-to-core noise correlations have also been exploited to achieve sub-femtosecond path stabilization for quantum networking at 100\% duty cycle~\cite{Nakamura26}.

A critical parameter limiting performance in all these applications is
\textit{inter-core skew} (ICS), the differential group delay between
signals co-propagating in different cores, arising from core-to-core
variations in effective refractive index, stress-induced birefringence,
and environmental perturbations. In SDM transmission, ICS forces digital
signal processing buffer sizes to scale with accumulated skew, degrading joint-core
carrier-phase estimation~\cite{alfredsson2019performance} and causing
cascaded penalties in multi-span links~\cite{cai2024impact}. In quantum
protocols relying on
interference of broadband photons, including QKD and entanglement
swapping, ICS can be
a critical decoherence source. ICS does not accumulate deterministically:
Lee~\textit{et al.} showed that azimuth-rotated splice attempts to cancel
ICS are unsuccessful~\cite{lee2017azimuth}, confirming that ICS is
dominated by stochastic, spatially-varying fluctuations and must be
{measured}, not engineered away. Existing classical techniques
such as C-OTDR~\cite{azendorf2020group,jin2022measurement} and
white-light interferometry~\cite{lee2015multi,lee2017azimuth} achieve
resolutions of only $10$--$20\,$ps---adequate for nanosecond-scale skew
in long fibers but far short of the sub-picosecond precision required for
quantum applications and short integrated-photonic segments.

Here we demonstrate that two-photon Hong-Ou-Mandel interference \cite{hong1987measurement} in a
fiber-integrated $4\times4$ multiport beam splitter provides ICS
measurement with a demonstrated precision of $\pm0.11\,$ps and a
fundamental Cram\'er--Rao precision limit in the femtosecond range. We
use photon pairs from spontaneous parametric down-conversion (SPDC)
through a commercial four-core fiber (Fibercore SM-4C1500(8.0/125)/001)
and a fiber-integrated MCF beam splitter~\cite{carine2020multi}. The
position of each HOM dip or peak directly encodes the differential group
delay between a core pair; Gaussian fits across all twelve input
combinations yield the complete ICS profile, and the root-mean-square
skew over lengths from $7.7\,$m to $1300\,$m provides the first direct
multi-length validation of the expected $\sigma_\tau \propto \sqrt{L}$
scaling. HOM interferometry requires no phase stabilization---the
coincidence dip depends only on the second-order temporal envelope,
immune to first-order path fluctuations~\cite{Aguilar2020}---making it
robust for measurements through long installed fibers where classical
interferometric methods would require active
stabilization~\cite{aboussouan2010high,gerrits2015spectral}. It is
precisely this phase-insensitivity that enabled the multi-length scaling
demonstration reported here: perturbation-sensitive classical methods
are impractical for long installed fibers, and prior ICS measurements
have not fitted the $\sqrt{L}$ dependence across a range of fiber
lengths~\cite{puttnam2019characteristics}.

\section{Inter-core skew: physical origin, scaling, and statistical
characterization}
\label{sec:ICS}

\begin{figure}
  \centering
  \includegraphics[width=\linewidth]{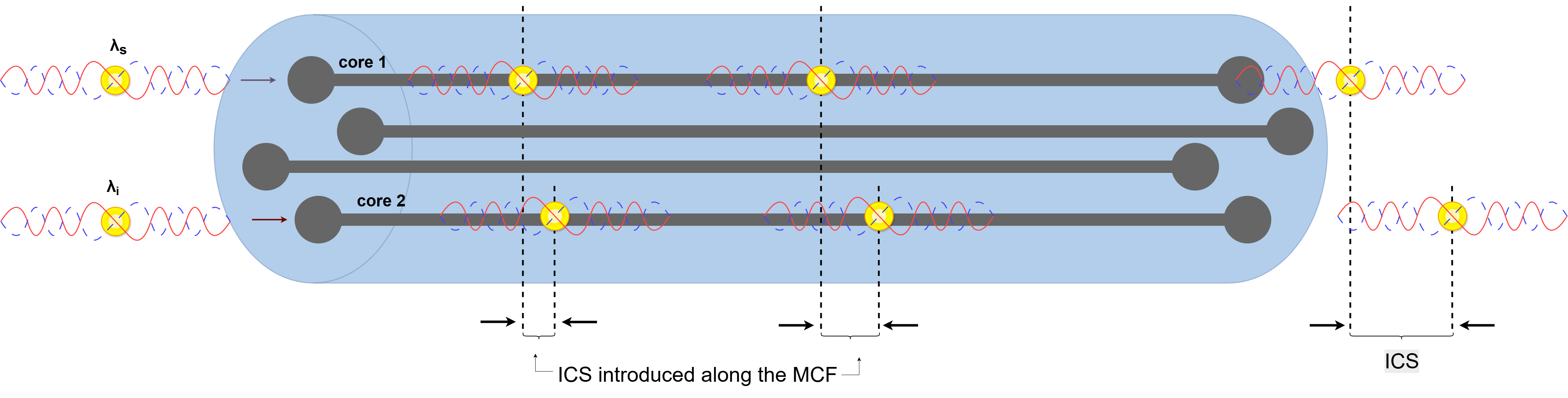}
  \caption{Illustration of ICS accumulation in a MCF.
  Identical pulses propagate (here from left to right) through different cores of the MCF. Because both
  pulses have the same frequency spectrum, any accumulated differential delay
  arises solely from core-dependent group velocity variations.}
  \label{fig:ICS}
\end{figure}

Inter-core skew (ICS), also referred to as differential group delay
(DGD) between cores, quantifies the temporal delay difference between optical signals
propagating through different cores of an MCF~\cite{puttnam2019characteristics}, as illustrated in
Fig.~\ref{fig:ICS}. ICS originates from core-to-core variations in
group velocity arising from small structural or refractive index
differences between cores, stress-induced birefringence, and
environmental perturbations such as bending or temperature
fluctuations~\cite{saleh2019fundamentals}.

A signal propagating through core $k$ of an MCF of length $L$ accumulates
an absolute group delay
$
  \tau_k(L) = {L\,n_{g,k}}/{c},
$
where $n_{g,k}$ is the group refractive
index of core $k$, and $c$ is the speed of light in vacuum.
The ICS between cores $j$ and $k$ is the \textit{difference} of these
absolute delays,
\begin{equation}
  \tau_{jk}(L)
  \equiv \tau_j(L) - \tau_k(L)
  = \frac{L}{c}\!\left(n_{g,j} - n_{g,k}\right).
  \label{eq:DGD}
\end{equation}
Note that $\tau_{jk} = -\tau_{kj}$; throughout
this paper we report the magnitude $|\tau_{jk}|$.

A purely deterministic refractive-index mismatch would produce an ICS
scaling linearly with $L$. Experiments consistently show that this
simple scaling is not observed~\cite{lee2017azimuth}. Rather, ICS is dominated
by stochastic, spatially-varying imperfections that produce local
positive and negative contributions to the differential delay. Lee
\textit{et al.}~\cite{lee2017azimuth} found that splicing rotations
designed to cancel ICS were unsuccessful, providing direct evidence of
the random character of ICS accumulation.

This behavior can be modeled by treating each absolute delay as a
one-dimensional zero-mean  random walk along the fiber.  The pairwise difference
$\tau_{jk} = \tau_j - \tau_k$ is then itself a zero-mean
random walk whose RMS grows as
\begin{equation}
  \sigma_{\tau}(L) = \kappa \sqrt{L},
  \label{eq:rms_scaling}
\end{equation}
with $\kappa$ the intrinsic ICS coefficient of the fiber (units:
$\mathrm{ps}/\sqrt{\mathrm{km}}$). This square-root scaling is formally analogous to the random-walk model
of polarization mode dispersion (PMD) in single-mode
fibers~\cite{poole1986phenomenological,poole1988statistical,saleh2019fundamentals},
and is consistent with ICS measurements reported in homogeneous
MCFs~\cite{puttnam2019characteristics}, though a direct multi-length
fit to the scaling law had not previously been demonstrated.

To go from pairwise delays to the RMS ICS, we use the fact
that a four-core fiber there are $\binom{4}{2} = 6$ unordered core pairs.
For each pair $(j,k)$ we perform measurements with both input orderings,
yielding two delay-stage positions $d_{jk}$ and $d_{kj}$ (12 ordered
measurements in total). Taking the half-difference
(Eq.~\ref{eq:measured_delay}) cancels the common arm imbalance
$\delta$ (see below) and gives the 6 unsigned ICS magnitudes
$|\tau_{jk}|$, one per unordered pair. Their RMS,
\begin{equation}
  \sigma_\tau(L)
  = \sqrt{\frac{1}{6}\sum_{\{j,k\}}|\tau_{jk}|^2},
  \label{eq:rms_def}
\end{equation}
characterizes the spread of delays across the core ensemble at length $L$. 
The measured RMS also includes a fixed, length-independent contribution
from single-mode fiber pigtails and fan-in/fan-out connectors at the
interferometer inputs. We therefore fit
\begin{equation}
  \sigma_\tau(L) = \kappa \sqrt{L} + c,
  \label{eq:rms_fit}
\end{equation}
where $c$ is a length-independent offset and $\kappa$ provides an
intrinsic, setup-independent measure of fiber-induced ICS.

To extract the ICS estimate, we must cancel out the offsets due to arm-length differences.
Let the two SMF interferometer input arms have fixed delays $\delta_\mathrm{I}$
and $\delta_\mathrm{II}$ arising from different SMF pigtail lengths, and
define the arm imbalance $\delta \equiv \delta_\mathrm{II} - \delta_\mathrm{I}$.
Let the delay stage position be $d$ (in arm~I). For input pair $(j,k)$,
the HOM dip appears at delay position $d_{jk}$ satisfying
\begin{equation}
  \delta_\mathrm{I} + d_{jk} + \tau_j = \delta_\mathrm{II} + \tau_k,
  \label{eq:dip_condition}
\end{equation}
giving $d_{jk} = \delta + \tau_{jk}$. Swapping
inputs gives $d_{kj} = \delta - \tau_{jk}$. Taking the
half-difference eliminates $\delta$:
\begin{equation}
  \tau_{jk} = \frac{d_{jk} - d_{kj}}{2}.
\label{eq:measured_delay}
\end{equation}
This is the ICS extracted for each of the 6 unordered core pairs using
both input orderings. The sum $d_{jk} + d_{kj} = 2\delta$
provides a consistency check on the arm imbalance.

\section{Multiport HOM interference in a \texorpdfstring{$4\times4$}{}
beam splitter}
\label{sec:HOM}

In the HOM experiment we use a $4 \times 4$ multiport beam splitter, which implements
a unitary transformation well-described by a normalized $4 \times 4$ real Hadamard matrix~\cite{carine2020multi}. 
 The input photon creation
operators $\hat{a}_{k}^\dagger$ transform to output operators
$\hat{b}_{l}^\dagger = \sum_k U_{lk}\hat{a}_{k}^\dagger$, where the $U_{lk}=\pm {1}/{2}$.

For two photons in input modes $i$ and $j$ with relative delay $\tau$,
the coincidence probability between output ports $k$ and $l$ is
\begin{equation}
  P_{kl}(\tau) = P_{kl}^{\rm dist}\bigl[1 - V_{kl}\,G(\tau)\bigr],
  \label{eq:Poftau}
\end{equation}
where $V_{kl}$ is the effective channel visibility and
$G(\tau) = \exp(-\tau^2/2\sigma_\tau^2)$ is the Gaussian envelope set
by the photon coherence time $\sigma_\tau$. The sign of $V_{kl}$ is
determined by the relative signs of the Hadamard matrix elements for
ports $k$ and $l$: channels for which $U_{ki}U_{lj}$ and $U_{li}U_{kj}$
share the same sign give destructive interference (bunching dip,
$V_{kl}>0$); opposite signs give constructive interference
(anti-bunching peak, $V_{kl}<0$). For a Fock-state input the ideal
visibility magnitude is $|V_{kl}|=1$ for both types.

In the presence of ICS, a differential group delay $\tau_{jk}$ between
the two input cores shifts the HOM feature center by exactly $\tau_{jk}$
along the delay axis without altering its shape or width. ICS therefore extracted from the {position} of the dip or peak,
making the method somewhat insensitive to loss imbalance, spectral
distinguishability, or multi-photon noise, which affect visibility but not
center position. Performing HOM measurements for all 12 input core-pair
combinations yields the complete differential delay profile of the fiber.

\section{Precision limits}
\label{sec:fisher}

In this experiment $\tau_{jk}$ is unknown and can be much larger than
the dip width $\sigma_\tau$, so the measurement is performed by scanning
the delay stage and locating the dip center by Gaussian fitting. The
precision limit of center estimation for a Gaussian profile is set by the
Cram\'{e}r--Rao bound (CRB)~\cite{jordan2022quantum}. For binary
coincidence trials, the Fisher information is
\begin{equation}
  F_{kl}(\tau)
  = \frac{1}{P_{kl}(\tau)\,[1 - P_{kl}(\tau)]}
    \left(\frac{dP_{kl}(\tau)}{d\tau}\right)^{\!2}.
  \label{eq:Fkm}
\end{equation}
Since the delay $\tau_{jk}$ is located by accumulating data over different delay positions, the total
Fisher information is obtained through the integral $\mathcal{I}_F =
\int_{-\infty}^{\infty} F_{kl}(u)\,du$, which holds since the scan range $\mathcal{R} \gg \sigma_\tau$ and the step size is much smaller than $\sigma_\tau$~\cite{jordan2022quantum,kay1993fundamentals}. Substituting the Gaussian
dip model Eq.~(\ref{eq:Poftau}), the scan CRB for locating the dip
center with $N$ total coincidence counts is~\cite{jordan2022quantum}
\begin{equation}
  \Delta\tau_{\rm CRB}
  = \frac{\sigma_\tau}{\sqrt{N_{\rm eff}\,J_F}},
  \label{eq:CRB_scan}
\end{equation}
where $N_{\rm eff} = N\,\sigma_\tau/\mathcal{R}$ is the effective count
density within one dip width 
and $J_F = \sigma_\tau\,\mathcal{I}_F$ is a dimensionless shape factor.
Narrower dips (broader photon spectrum) improve precision by reducing
$\sigma_\tau$, and $J_F \propto V^2$, so higher visibility
monotonically improves precision. For the SPDC measurements reported here,
the scan range is $\mathcal{R} = 1.8\,$ps and coincidence counts per scan
range from $N \approx 1.5\times10^5$ to $3.5\times10^5$, giving
effective count densities $N/\mathcal{R} \approx 8\times10^4$ to
$2\times10^5\,\mathrm{ps}^{-1}$. Summing over all six output pairs with
$\bar{V} \approx 0.92$ and $\sigma_\tau \approx 0.25\,$ps, the
per-channel Cram\'{e}r--Rao bound ranges from $0.93$ to $1.94\,$fs.
The conservative precision estimate of $\pm0.11\,$ps reported below is thus about two
orders of magnitude above the fundamental quantum limit, which could
be approached with more precise delay control. We note that the
demonstrated $\pm0.11\,$ps precision is more than adequate for the
random-walk scaling measurement reported in
Section~\ref{sec:results}; the Fisher information analysis motivates
future experiments targeting the femtosecond regime.

\section{Experimental implementation}
\label{sec:experiment}
\begin{figure}
  \centering
  \includegraphics[width=\linewidth]{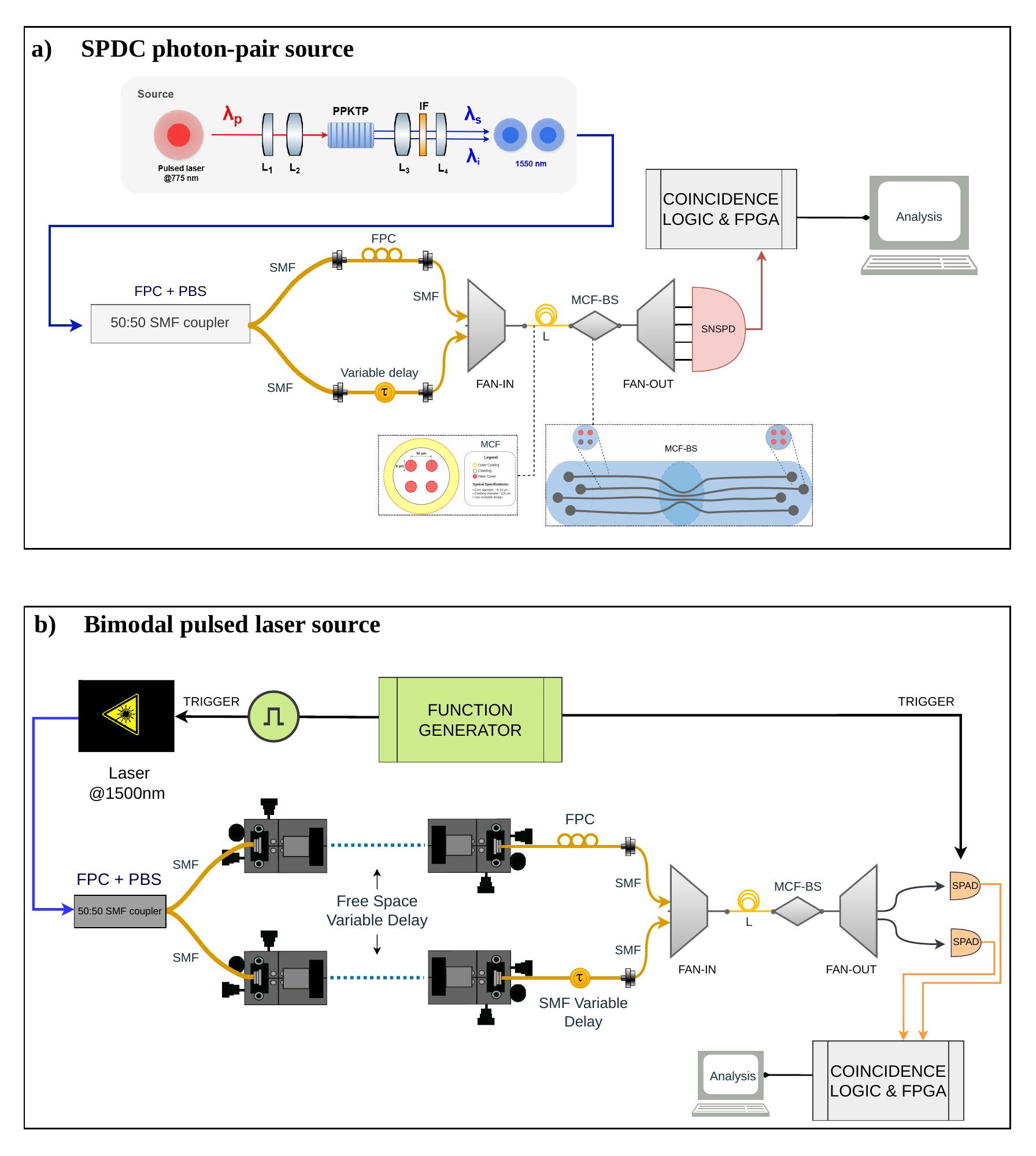}
  \caption{a) Experimental setup for HOM interference using SPDC source. The 
  photons are coupled into SMFs, separated by a variable delay $\tau$,
  and injected into the four-core MCF via a FAN-IN. Coincidence counts are registered
  by SNSPDs and an FPGA. b) Experimental setup for HOM interference using a weak laser source. Additional optical delay lines were added.  See text for additional details.}
  \label{fig:setup}
\end{figure}

\subsection{Photon-pair source and interferometer}
The experimental setup is shown in Fig.~\ref{fig:setup}.
Photon pairs were generated via type-II SPDC in a periodically poled
KTiOPO$_4$ (PPKTP) crystal pumped by a mode-locked laser at $775\,$nm
with repetition rate $f_{\rm rep} = 80\,$MHz and $\sim\!70\,$mW
average power. Degenerate photon pairs centered at $1550\,$nm were separated by
a polarizing beam splitter (PBS) and spectrally filtered by a $30\,$nm
interference filter (IF) before coupling into single-mode fibers
(SMFs). 

The photons were routed through a modified Mach--Zehnder interferometer
(TPI) with a variable free-space delay line providing up to $15\,$mm
path difference ($\approx 50\,$ps). Both photons were injected into the
MCF of length $L$ via the MUX, and then sent to interfere in the MCF-BS.
To study ICS as a function of length, the measurement was repeated for
six separate fiber segments ($L = 7.7, 12.7, 49.7, 72.7, 114.7,$ and
$187.7\,$m); each segment is an independently handled cut of
Fibercore SM-4C1500(8.0/125)/001 four-core fiber~\cite{fibercore2024datasheet}
with a square core arrangement, $50\,\mu\mathrm{m}$ nominal core spacing,
$125\,\mu\mathrm{m}$ cladding diameter, and mode field diameter of
$7.4$--$8.5\,\mu\mathrm{m}$ at $1550\,$nm (single-mode operating window
$1520$--$1650\,$nm).
Inter-core crosstalk during propagation is negligible, so that mode
mixing occurs only in the engineered beam-splitter coupling region.

The multiport beam splitter is a $4\times4$ MCF-BS implementing the
unitary Hadamard transformation via evanescent coupling
between the four cores over a finite interaction
length, as described and characterized in Refs. ~\cite{carine2020multi,carine2021maximizing}. Input and output
cores are interfaced via fiber-optic MUX and DEMUX fan-in/fan-out
devices, as shown in Fig.~\ref{fig:setup}.
\par
The four output modes were detected by
superconducting nanowire single-photon detectors (SNSPDs). Coincidence
events were registered by an FPGA counting system.  For the pulsed SPDC source, accidental coincidences accumulate at the
rate
$R_{\rm acc} = ({R_A \cdot R_B})/{f_{\rm rep}}$,
where $R_A$ and $R_B$ are the single-channel detection rates.  Accidentals
have 
been subtracted from all SPDC data.

The $\approx 100\,$ps delay range sets a practical upper bound on
observable ICS: core pairs whose differential delays exceed this limit
will not show an HOM feature within the scan. This explains the
progressive reduction in visible dip/peak combinations at longer fiber
lengths (Section~\ref{sec:results}). The delay line was extended
to $>10\,$cm ($>670\,$ps) to enable measurements at the longer
installed fiber lengths at the cost of increased losses; results are reported in
Section~\ref{sec:results}.

\subsection{Bimodal pulsed laser source for long installed fibers}
\label{sec:bimodal}
For the field-installed $1300\,$m fiber segment, ICS is still measured
via HOM two-photon interference, but using a pulsed laser source in
place of SPDC. This substitution is necessary because the SPDC source
becomes impractical at this length due to channel losses, which are
large and dominated not by fiber propagation but by MCF-to-MCF
connectors and splices.
The bulk of the  $10.70\,$dB of channel loss arises from butt-coupled MCF connectors
(individual insertion loss $0.3$--$3.1\,$dB, requiring manual
rotational alignment of the square core pattern before each
connection), fusion splices, and the DEMUX fan-out. 
For SPDC, both photons must survive their respective arms
independently, so a per-arm loss of $\sim 11\,$dB reduces
coincidence rates by a factor of $\sim\!160$, making
measurements impractical.

The setup is shown schematically in Fig.~\ref{fig:setup}(b). A single $\sim\!300\,$ps telecom laser
pulse is split at the SMF coupler into two arms; both arms traverse
the MCF under test (one per core), accumulate the inter-core delay
$\tau_{jk}$, and recombine at a second SMF coupler acting as the
HOM beam splitter. The two outputs are detected by InGaAs SPADs triggered by
the function generator driving the laser at $f_{\rm rep} = 1\,$MHz.

The laser has a multi-mode longitudinal mode structure visible as a
sinusoidal beating of the fast single count oscillations along the delay axis. The
beat period is $T_{\rm beat} \approx 10\,$ps, corresponding to a mode
spacing $\delta\nu \approx 100\,$GHz ($\approx 0.8\,$nm at $1550\,$nm).

\textbf{Two-photon decomposition and the 50\% visibility ceiling.}
The physics of this measurement is most clearly understood by
considering the two-photon component of the coherent pulse, which
dominates the coincidence signal. When a two-photon pulse is split at
the input $50:50$ coupler, two outcomes are equally likely: (i) one
photon takes each arm (probability $\frac{1}{2}$), leading to
standard HOM interference at the output coupler; (ii) both photons
take the same arm (probability $\frac{1}{2}$), forming a
$|2,0\rangle$ or $|0,2\rangle$ state that produces N00N-state
interference at the output coupler with fringe period $\lambda/2
\approx 775\,$nm in path length. The coincidence probability is a combination of
both contributions:
$
  P_{\rm coinc}(\tau) =
  \frac{1}{2}{P_{\rm HOM}(\tau)}
  +
  \frac{1}{2}{P_{\rm N00N}(\tau)}.$
The N00N fringes oscillate at period $\lambda/2 \approx 775\,$nm,
sensitive to sub-wavelength optical path fluctuations. Ambient thermal
and mechanical drifts during the $\approx\!30\,$s integration at each
delay point cause path-length excursions of several micrometers, sweeping
through multiple fringe cycles and averaging the N00N contribution to its
mean: $\langle P_{\rm N00N}\rangle = \frac{1}{2}$. The HOM envelope
width $\sigma_\tau c \approx 120\,\mu$m is four orders of magnitude
larger than $\lambda/2$, so the same drift is entirely negligible for
the HOM term. The measured
coincidence rate therefore becomes
\begin{equation}
  R_c(\tau) = R_c^{\infty}\!\left[
    \frac{1}{2} + \frac{1}{2}\left(1 - G_{\rm HOM}(\tau)\right)
  \right]
  = R_c^{\infty}\!\left[1 - \frac{1}{2}\,G_{\rm HOM}(\tau)\right],
  \label{eq:bimodal_pattern}
\end{equation}
where $G_{\rm HOM}(\tau)$ is the HOM envelope (Gaussian modulated by
the bimodal beats). The maximum visibility is therefore
$V_{\rm max} = 50\%$, and is
unchanged by the bimodal spectrum. The bimodal beats enter through
$G_{\rm HOM}(\tau)$:
\begin{equation}
  G_{\rm HOM}(\tau) =
  \cos^2\!\!\left(\frac{\pi\tau}{T_{\rm beat}}\right)
  \exp\!\left(-\frac{\tau^2}{2\sigma_{\rm pulse}^2}\right),
  \label{eq:GHOM}
\end{equation}
where $\sigma_{\rm pulse} \approx 127\,$ps for a transform-limited
$\sim\!300\,$ps pulse. The ICS between a core pair shifts the entire
pattern by $\tau_{jk}$. In practice, a coarse delay scan is first
performed to identify the beat period at which fringe visibility is
maximum, locating $\tau_{jk}$ to within $\sim\!T_{\rm beat}$. A
fine-step scan is then performed over a single beat period centered on
this maximum, and a fit to the $\cos^2$ envelope of
Eq.~(\ref{eq:GHOM}) extracts the fringe-center position, giving
$\tau_{jk}$ with an effective precision
$\sigma_{\rm eff} \approx T_{\rm beat}/2\pi \approx 1.6\,$ps,
far below the $\sim\!300\,$ps pulse width. 
\par
The laser power is adjusted so that link attenuation reduces the mean
photon number to $\mu_{\rm eff} \approx 0.5$ photons/pulse per arm at
the detectors, the regime where the two-photon decomposition above
applies.
This corresponds to 
$R_A \approx R_B \approx 50\,000\,$cps per detector and
$R_{\rm coinc} \approx 2600\,$cps, {regardless of which link
is under test}. 
 We note that here the same connector losses that render SPDC impractical thus
serve as a natural attenuator.  Dark-count
accidentals ($\lesssim\!0.01\,$cps) were negligible.

We therefore present two complementary measurement regimes: (A)~short
laboratory fibers ($7.7$--$187.7\,$m) using SPDC photon pairs and  (B)~long field-deployed fibers ($1300\,$m) using the
 pulsed laser.

\section{Results and discussion}
\label{sec:results}

\subsection{Multiport HOM interference, visibilities, and dip-center extraction}
\label{sec:igdextract}

\begin{figure}
  \centering
  \begin{subfigure}{0.48\linewidth}
    \centering
    \includegraphics[width=\linewidth]{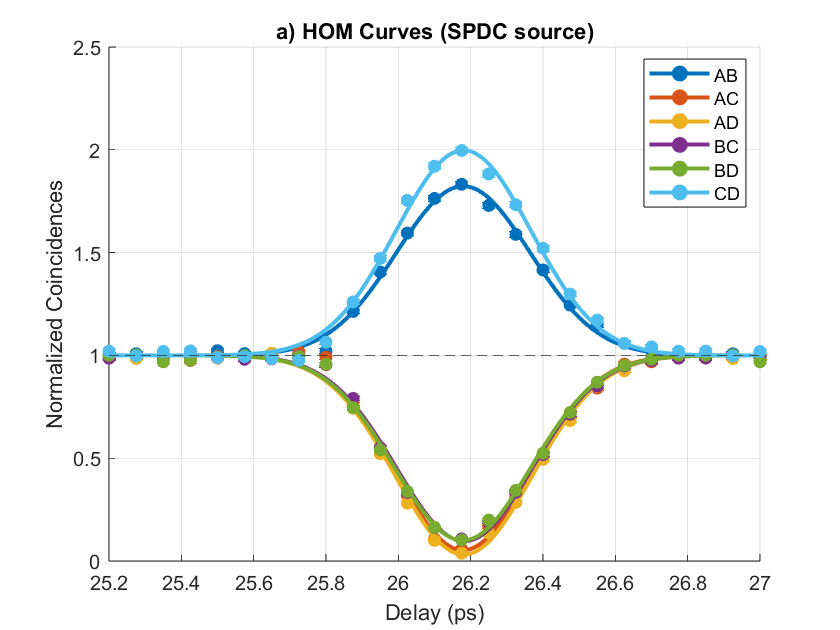}
  \end{subfigure}
  \hfill
  \begin{subfigure}{0.48\linewidth}
    \centering
    \includegraphics[width=\linewidth]{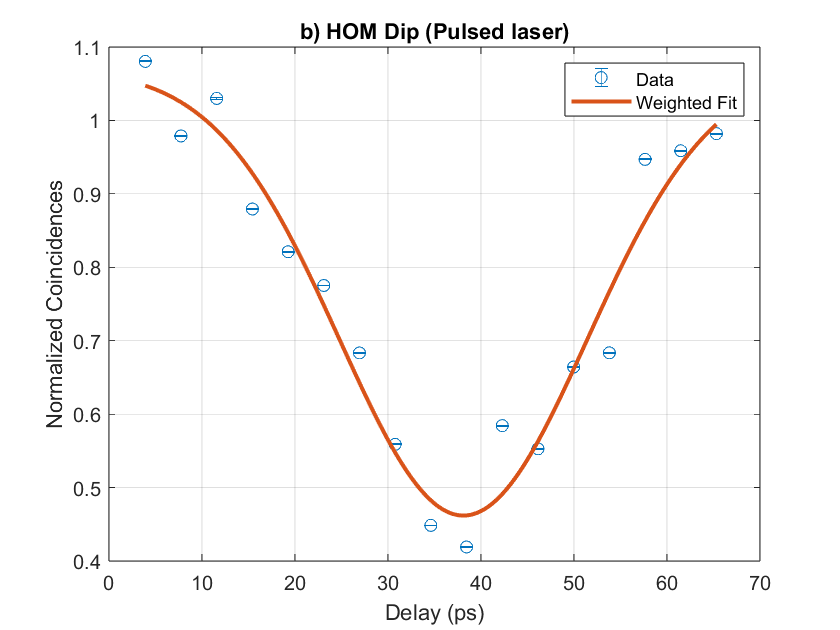}
  \end{subfigure}

  \caption{HOM interference data.
  (a)~Normalised coincidence counts
  for all six output pairs as a function of relative delay for a
  $12.7\,$m MCF, input combination $a_3$,$a_4$, using the SPDC source.
  Solid lines are Gaussian fits. Statistical error bars are smaller than
  the data points and not visible at this scale. Coincidence curves show bunching dips
  (AC, AD, BC, BD) and anti-bunching peaks (AB, CD) as predicted by
  the $4\times4$ Hadamard transformation. Fit parameters and
  visibilities are listed in Table~\ref{tab:visibilities_and_fits}.
  (b)~Coincidence rate as a function of relative delay for the
  $1300\,$m installed fiber, input combination $a_3$,$a_4$ for a single pair of outputs, using
  the bimodal pulsed laser source. The modulated envelope reflects
  the $\cos^2$ beat structure of Eq.~(\ref{eq:GHOM}); the solid line
  is a fit used to extract $\tau_{jk}$.}
  \label{fig:Coincidences_5m}
\end{figure}

As an example of the extracted data using HOM interference of SPDC photon pairs, Fig.~\ref{fig:Coincidences_5m}(a) shows
coincidence rates for a $L = 12.7\,$m MCF segment and one
input core-pair combination ($a_3, a_4$) as a function of relative
delay $\tau$. The coincidence curves display the
characteristic multiport HOM landscape: four output pairs (AC, AD, BC,
BD) show bunching dips and two pairs (AB, CD) show anti-bunching peaks,
in quantitative agreement with the predictions for the Hadamard
transformation described in Section~\ref{sec:HOM}.
The dip-center positions are extracted by fitting each coincidence curve
to a Gaussian,
\begin{equation}
  y = a + b\exp\!\left(-\frac{(x - d)^2}{2 \sigma_\tau^2}\right),
  \label{eq:gaussian_fit}
\end{equation}
where $d$ gives the dip-center delay, $\sigma_\tau$ is the width,
$b$ is the amplitude, and $a$ is the baseline. The visibility is
$V = |b|/a$, and the center position $d$ is insensitive to visibility
degradation from loss or spectral effects.

\begin{table}
  \caption{Measured Gaussian fit
  parameters of HOM interference for all six output channel pairs, for inputs 3--4 in the $L=12.7\,$m fiber.
  Uncertainties are statistical fit uncertainties $\sigma_{\rm fit}$ only;
  the delay-stage positioning uncertainty $\sigma_{\rm cal}$ is not included here
  (see text).}
  \label{tab:visibilities_and_fits}
  \centering
  \small
  \begin{tabular}{c|c|c|c|c}
    \hline
    Channel & Type & Visibility & $d$ (ps) & $\sigma_\tau$ (ps) \\
    \hline
    AB & peak & $0.823 \pm 0.018$ & $0.262 \pm 0.007$ & $0.247 \pm 0.006$ \\
    AC & dip  & $0.947 \pm 0.023$ & $0.261 \pm 0.007$ & $0.247 \pm 0.007$ \\
    AD & dip  & $0.968 \pm 0.019$ & $0.267 \pm 0.006$ & $0.252 \pm 0.006$ \\
    BC & dip  & $0.903 \pm 0.017$ & $0.265 \pm 0.006$ & $0.251 \pm 0.006$ \\
    BD & dip  & $0.900 \pm 0.018$ & $0.266 \pm 0.006$ & $0.251 \pm 0.006$ \\
    CD & peak & $0.998 \pm 0.021$ & $0.264 \pm 0.007$ & $0.250 \pm 0.006$ \\
    \hline
    Mean & & $0.92 \pm 0.06$ & $0.264 \pm 0.001$ & $0.250 \pm 0.002$ \\
    \hline
  \end{tabular}
\end{table}

The fit data is shown in Table \ref{tab:visibilities_and_fits}. The visibilities range from $82\%$ to near $100\%$, with a mean
of $\bar{V} = 92 \pm 6\%$ after accidental subtraction, close to the
ideal $|V|=1$ expected for a Fock-state input.  Deviations from ideal arise from
residual photon distinguishability, fan-in/fan-out asymmetries, and
imperfect beam-splitter split ratios.
 The mean coherence time $\sigma_\tau = 0.250 \pm 0.002\,$ps confirms
the expected value from the $30\,$nm spectral filter. Gaussian fits
of all six output channels yield consistent dip-center positions
($\bar{d} = 0.264 \pm 0.001\,$ps), as expected since the dip position
depends only on the input core pair (Table~\ref{tab:visibilities_and_fits}). 

We now quantify the uncertainty on these extracted values.
The $\pm0.002\,$ps fit precision $\sigma_{\rm fit}$ is the purely
statistical uncertainty from averaging over six output channels.
The dominant contribution to the dip-position uncertainty is the
delay-stage positioning uncertainty $\sigma_{\rm cal} = 0.15\,$ps,
taken as a conservative upper bound from the manufacturer-specified
delay stability of the General Photonics VDL-002 delay line.
These two contributions combine as
$\sigma_d = (\sigma_{\rm fit}^2 + \sigma_{\rm cal}^2)^{1/2}
\approx 0.15\,$ps for each dip position, as listed in
Table~\ref{tab:IGDs}. Since the ICS is extracted via the
half-difference of two independent readings
(Eq.~\ref{eq:measured_delay}), the propagated uncertainty on each
$\tau_{jk}$ is $\sigma_{\tau_{jk}} = \sigma_d/\sqrt{2} \approx 0.11\,$ps.
The demonstrated precision of $\pm0.11\,$ps is thus a consequence of
the delay stage rather than photon statistics, and could be
substantially improved toward the femtosecond CRB with
interferometrically calibrated delay control.

\begin{table}[htbp]
  \caption{Extracted HOM dip positions $d_{ij}$ for all 12
  ordered input channel pairs of the MCF at $L=12.7\,$m and $L=49.7\,$m.
  Uncertainties reflect $\sigma_d = (\sigma_{\rm fit}^2 +
  \sigma_{\rm cal}^2)^{1/2} \approx 0.15\,$ps, dominated by the
  delay-stage positioning uncertainty. The ICS for each unordered pair
  is obtained as $\tau_{ij} = |d_{ij} - d_{ji}|/2$.
  }
  \label{tab:IGDs}
  \centering
  \begin{tabular}{c|c|c}
    \hline
    Input pair & $d_{ij}$ (ps), $L=12.7\,$m & $d_{ij}$ (ps), $L=49.7\,$m \\
    \hline
    1--2 & $13.73 \pm 0.15$ & $31.57 \pm 0.15$ \\
    1--3 & $16.22 \pm 0.15$ & $12.67 \pm 0.15$ \\
    1--4 & $23.88 \pm 0.15$ & $ 6.62 \pm 0.15$ \\
    2--1 & $23.98 \pm 0.15$ & $ 7.31 \pm 0.15$ \\
    2--3 & $21.40 \pm 0.15$ & $32.67 \pm 0.15$ \\
    2--4 & $29.06 \pm 0.15$ & $60.51 \pm 0.15$ \\
    3--1 & $21.62 \pm 0.15$ & $27.42 \pm 0.15$ \\
    3--2 & $16.49 \pm 0.15$ & $38.63 \pm 0.15$ \\
    3--4 & $26.53 \pm 0.15$ & $17.15 \pm 0.15$ \\
    4--1 & $14.15 \pm 0.15$ & $31.07 \pm 0.15$ \\
    4--2 & $ 8.88 \pm 0.15$ & $41.51 \pm 0.15$ \\
    4--3 & $11.30 \pm 0.15$ & $22.64 \pm 0.15$ \\
    \hline
  \end{tabular}
\end{table}

The extracted dip positions $d_{ij}$ for all 12 input combinations for fibers with $L = 12.7\,$m and $L = 49.7\,$m
are listed in Table~\ref{tab:IGDs}. Each pair is measured in both
input orderings; the half-difference (Eq.~\ref{eq:measured_delay})
yields the ICS $\tau_{jk}$, given in Table~\ref{tab:DGD_values}, along with propagated uncertainties and
RMS values. The RMS ICS increases from $12.20 \pm 0.045\,$ps at $12.7\,$m
to $17.47 \pm 0.045\,$ps at $49.7\,$m, reflecting the cumulative
accumulation of inter-core skew with propagation length.

\begin{table}[htbp]
  \caption{Inter-core skew (ICS) magnitudes $\tau_{ij}=|d_{ij}-d_{ji}|/2$
  for each unordered core pair at $L=12.7\,$m and $L=49.7\,$m.
  Individual ICS uncertainties $\sigma_{\tau_{jk}} = \sigma_d/\sqrt{2}
  \approx 0.11\,$ps; RMS uncertainty
  $\sigma_{\rm RMS} = \sigma_{\tau_{jk}}/\sqrt{6} \approx 0.045\,$ps.}
  \label{tab:DGD_values}
  \centering
  \begin{tabular}{c|c|c}
    \hline
    Pair $i$-$j$ & $|\tau_{ij}|$\,(ps), $L=12.7\,$m & $|\tau_{ij}|$\,(ps), $L=49.7\,$m \\
    \hline
    1--2 & $10.25 \pm 0.11$ & $24.26 \pm 0.11$ \\
    1--3 & $ 5.41 \pm 0.11$ & $14.75 \pm 0.11$ \\
    1--4 & $ 9.73 \pm 0.11$ & $24.45 \pm 0.11$ \\
    2--3 & $ 4.91 \pm 0.11$ & $ 5.97 \pm 0.11$ \\
    2--4 & $20.18 \pm 0.11$ & $18.99 \pm 0.11$ \\
    3--4 & $15.24 \pm 0.11$ & $ 5.49 \pm 0.11$ \\
    \hline
    RMS  & $12.20 \pm 0.045$ & $17.47 \pm 0.045$ \\
    \hline
  \end{tabular}
\end{table}


\subsection{Random-walk scaling of ICS with fiber length}
\label{sec:scaling}

\begin{figure}
  \centering
  \includegraphics[width=\linewidth]{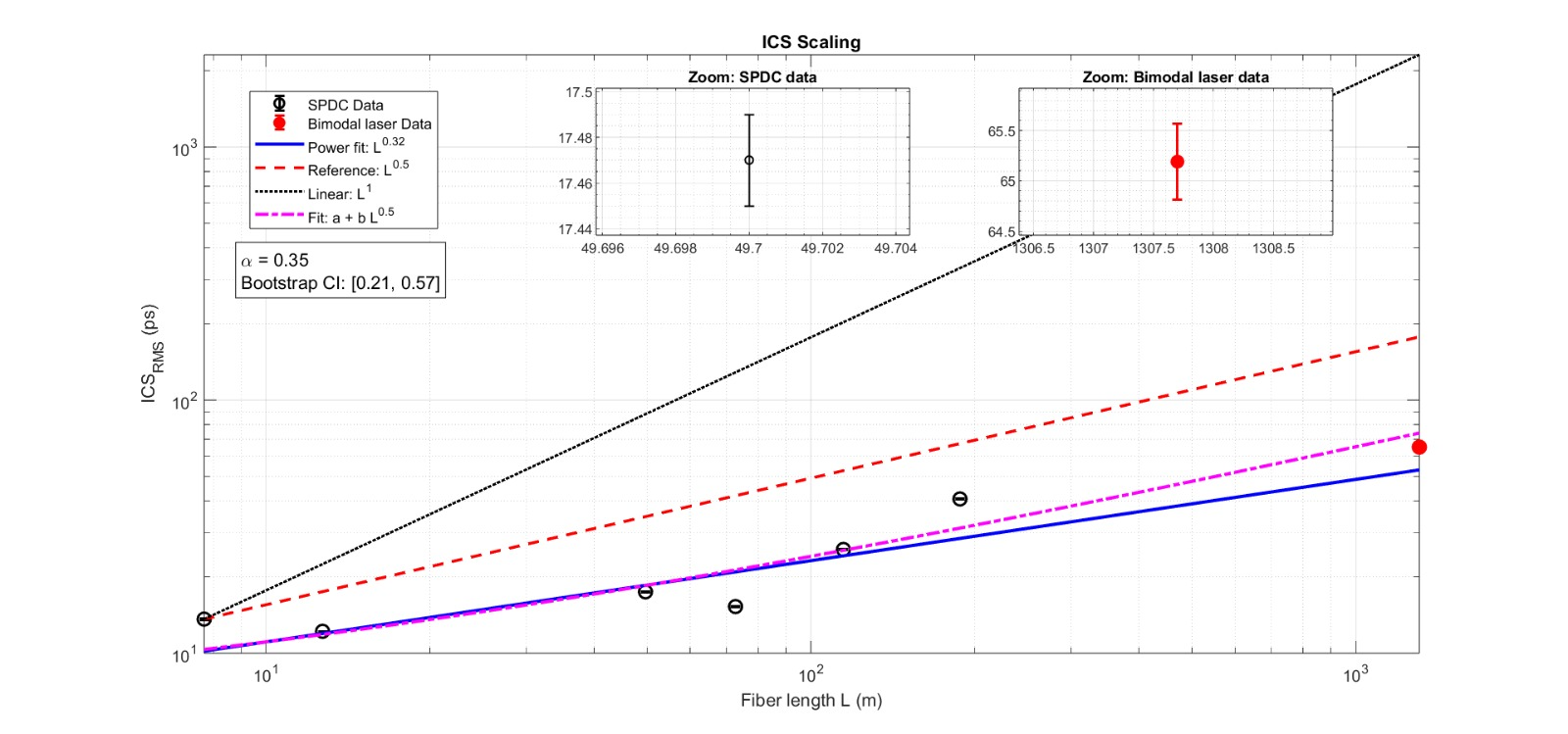}
  \caption{Scaling of the ICS$_{\rm RMS}$ as a function of fiber
  length $L$ in log-log scale for all measured lengths
  $L$. Experimental measurements
  (markers with error bars) are fitted using Eq. (\ref{eq:rms_fit}) (dot-dashed pink line) and a weighted power-law model
  $\mathrm{ICS} \propto L^{\alpha}$ (solid blue line). The shaded
  region represents the propagated uncertainty of the fit in log space.
  Reference scalings $L^{0.5}$ (dashed red) and $L^{1}$ (dotted black)
  are shown for comparison. A bootstrap resampling analysis
  ($N=5000$) yields a 95\% confidence interval for $\alpha=0.35$ with a 95\% CI [0.21,0.57].The
  point estimate $\alpha$ is biased below 0.5 by the physical
  offset $c$ in the underlying model (see text). Data at
  $1300\,$m were obtained using the bimodal pulsed laser source
  (regime~B; Section~\ref{sec:bimodal}).}
  \label{fig:DGD_scaling_bootstrap}
\end{figure}

Figure~\ref{fig:DGD_scaling_bootstrap} shows the RMS ICS $\sigma_\tau(L)$ as a function
of $\sqrt{L}$ for fiber lengths $L =$ 7.7, 12.7, 49.7, 72.7, 114.7, 187.7 m measured
with SPDC (regime~A, Section~\ref{sec:bimodal}).  At longer lengths, some core-pair combinations no longer exhibit HOM
features within the original delay scan range ($\approx\!100\,$ps):
10 of 12 input pairs were observable at $L = 72.7\,$m, 7 at
$L = 114.7\,$m, and only 1 on an installed $L = 187.7\,$m fiber (not shown).
This is a direct consequence of ICS exceeding the TPI delay-line
range and is consistent with the $\kappa\sqrt{L}$ scaling. To access
these longer lengths it was necessary to extend the free-space delay
line, adding approximately $-6\,$dB of coupling loss per photon and
providing a further practical motivation for the pulsed-laser
approach, which was used to test a field-deployed $L\approx 1300$\,m fiber (red data point). The shortest point at $L = 7.7\,$m
corresponds to the MCF fiber contained within the MUX fan-in and MCF-BS devices. The data follow the
 Eq.~(\ref{eq:rms_fit}), confirming the
stochastic random-walk model. A weighted linear fit gives
\begin{equation*}
  \kappa = 1.54 \pm 0.08\;\mathrm{ps}/\!\sqrt{\mathrm{m}}
         = 48.7 \pm 2.5\;\mathrm{ps}/\!\sqrt{\mathrm{km}},
  \qquad
  c = 9.76 \pm 1.2\;\mathrm{ps}.
\end{equation*}
The value $\kappa = 48.7\,\mathrm{ps}/\!\sqrt{\mathrm{km}}$ is 
$> 50$ larger than typical PMD coefficients for single-mode
fiber ($0.1$--$1\,\mathrm{ps}/\sqrt{\mathrm{km}}$), confirming that the
measured effect is genuine inter-core differential group delay rather
than a polarization-related artifact. Expressed as a skew per unit length,
$\kappa = 48.7 \pm 2.5\,\mathrm{ps}/\!\sqrt{\mathrm{km}}$ corresponds
to $0.049 \pm 0.003\,\mathrm{ns/km}$ at $L = 1\,$km, consistent with
reported ICS values of $0.06$--$0.2\,\mathrm{ns/km}$ for other
four-core and seven-core fiber
designs~\cite{puttnam2018inter,azendorf2020group}.
The length-independent intercept $c = 9.76\,$ps  is attributed to step-function ICS offsets at each butt-coupled
MCF connector junction, where independently manufactured fiber sections
present a fixed, random differential group-index mismatch. 

While $\sqrt{L}$ scaling is expected for purely stochastic
accumulation, finite sample size or correlated perturbations may cause
deviations, so we also fit a general power-law model
$\sigma_\tau(L) = A\,L^{\alpha}$ (Eq.~\ref{eq:general_scaling}), which
reduces to the stochastic model for $\alpha = 0.5$ and to deterministic
linear scaling for $\alpha = 1$. A weighted log-log regression
(Appendix~\ref{app:inference}) yields a point estimate $\alpha=0.32$, which should be interpreted with care: the power-law model $A\,L^\alpha$ contains no intercept, so when the true behaviour includes a physical offset $c \neq 0$ (as here), the exponent is biased below its true value. This is precisely why the constrained $\sqrt{L}+c$ fit is
visually superior despite having the same number of free parameters.
The data are well described by both models, with the power-law fit yielding a higher goodness of fit ($R^2 \approx 0.97$) compared to the $\sqrt{L}+c$ model ($R^2 \approx 0.93$).
A bootstrap resampling analysis ($N = 5000$) gives $\alpha=0.35$ a 95\% confidence
interval $[0.21,0.57]$ that includes the theoretical
$\alpha = 0.5$. Thus, the null hypothesis of standard $\sqrt{L}$ scaling
cannot be rejected at the 95\% level.


\subsection{Comparison with classical ICS measurement methods}

Table~\ref{tab:comparison} contextualizes our results within the
existing literature on ICS characterization. The key comparison is in
precision: C-OTDR achieves $\lesssim\!20\,$ps resolution after Gaussian
fitting of the correlation peak~\cite{azendorf2020group}, while our HOM
method achieves $\pm0.11\,$ps from dip-center fitting, a factor of
$\sim\!180$ improvement in demonstrated precision. The fundamental CRB
for the measurements reported here (per-channel values of $0.93$--$1.94\,$fs;
Section~\ref{sec:fisher}) is more than four orders of magnitude below
the demonstrated precision, indicating that substantially greater
precision is in principle achievable with an improved delay calibration.

\begin{table}[htbp]
  \caption{Comparison of ICS characterization methods.
  {Skew} column definitions vary by method. C-OTDR and
  cross-correlation entries give an average or maximum
  $|\tau_{jk}|/L$ per core pair. The HOM/SPDC entry gives the
  random-walk coefficient $\kappa$ (units
$\mathrm{ps}/\!\sqrt{\mathrm{km}}$) from fitting the RMS
  over all six unordered pairs. The HOM/laser entry gives the mean
  $|\tau_{jk}|/L$ averaged over all six unordered pairs (see
  note~$^e$). Direct numerical comparison across rows is therefore
  qualitative. {Precision}: demonstrated single-measurement
  resolution for each method.
  C-OTDR: correlation OTDR. WLI: white-light interferometry.}
  \label{tab:comparison}
  \centering
  \small
  \begin{tabular}{llllll}
    \hline
    Method & Cores & $L$ & Skew & Precision & Ref.\\
    \hline
    C-OTDR & 7  & 10\,km & 0.18\,ns/km & $\lesssim\!20$\,ps & \cite{azendorf2020group}\\
    C-OTDR & 7  &  1\,km & 1.2\,ns/km  & $\lesssim\!20$\,ps & \cite{azendorf2020group}\\
    C-OTDR & 19 &  5\,km & 0.66\,ns/km & $\lesssim\!20$\,ps & \cite{azendorf2020group}\\
    C-OTDR & 19 & 25\,km & 1.2\,ns/km  & $\lesssim\!20$\,ps & \cite{azendorf2020group}\\
    MPS$^a$ & 19 & 5\,km & $\pm300$\,ps (drift) & $\sim$ns & \cite{azendorf2020group}\\
    WLI     & 4  & short & ---$^b$ & sub-ps  & \cite{lee2015multi}\\
    WLI+splice$^b$ & 4 & --- & ---$^b$ & sub-ps  & \cite{lee2017azimuth}\\
    Cross-corr.$^c$ & 7 & 53.7\,km & $\lesssim\!0.5$\,ns/km & sub-ps & \cite{puttnam2019characteristics}\\
    Temp.\ OTDR & 19 & --- & $\pm60$\,ps/40\,K & $\sim$ps & \cite{puttnam2018impact}\\
    \hline
    \textbf{HOM/SPDC}  & \textbf{4} & \textbf{7.7--187.7\,m} &
      $\boldsymbol{\kappa=48.7\pm2.5}$\,ps/$\sqrt{\text{km}}$ &
      $\boldsymbol{\pm0.11}$\,ps; CRB\,$\sim$fs &
      \textbf{here}\\
    \textbf{HOM/laser}$^e$  & \textbf{4} & \textbf{1.3\,km} &
      \textbf{0.049 ns/km}\, (mean) &
      \textbf{0.70\,ps} &
      \textbf{here}\\
    \hline
  \end{tabular}
  \par\smallskip
  \raggedright\footnotesize
  $^a$MPS: modulation phase shift. Only one core can be measured at a
  time, so consecutive measurements are separated by thermal drift of
  several hundred ps, preventing reliable differential skew
  characterisation.\\
  $^b$WLI references characterize skew at fixed lengths or demonstrate
  stochastic accumulation qualitatively; neither $\kappa$ nor absolute
  skew magnitudes were reported.\\
  $^c$Cross-correlation of 10\,Gb/s OOK waveforms~\cite{puttnam2019characteristics}.
  Skew entry is the maximum observed $|\tau_{jk}|/L$ (between the most
  disparate core pair); the mean over cores relative to the centre core
  is $\sim\!100$\,ps/km for homogeneous MCFs. Precision applies to
  dynamic skew tracking; static absolute delays are determined to
  ns-level accuracy.\\
  $^d$Rotated-splice compensation fails, confirming that ICS
  accumulation is stochastic~\cite{lee2017azimuth}.\\
  $^e$HOM/laser skew is the mean $|\tau_{jk}|/L$ averaged over all
  six unordered core pairs at $L=1300$\,m, the metric most directly
  comparable to literature average-skew figures. 
  \\
  Fiber: Fibercore SM-4C1500(8.0/125)/001 for this work.
\end{table}

\section{Conclusion}
\label{sec:conclusion}

We have demonstrated sub-picosecond measurement of inter-core skew
(differential group delay) in a commercially available four-core
multicore fiber (Fibercore SM-4C1500) using Hong--Ou--Mandel
two-photon interference in a fiber-integrated $4\times4$ multiport
beam splitter. ICS is extracted from HOM dip and peak center positions
across all twelve input core-pair combinations, with a demonstrated
precision of $\pm0.11\,$ps per measurement, limited by the delay-stage
positioning uncertainty, and a fundamental Cram\'er--Rao bound in the
femtosecond range that could be approached with improved  delay control.

The RMS ICS grows as
$\sigma_\tau(L)
= (48.7 \pm 2.5\,\mathrm{ps}/\!\sqrt{\mathrm{km}})\sqrt{L}
+ (9.76 \pm 1.2)\,$ps
over lengths from $7.7\,$m to $187.7\,$m, providing the first direct
validation of the stochastic random-walk scaling of ICS in a four-core
fiber across a length range inaccessible to classical methods.
Extension of the scaling curve to $L = 1300\,$m using the bimodal
pulsed laser source is consistent with the $\sqrt{L}$ model, though
a generalised power-law fit cannot conclusively rule out sub-diffusive scaling with
the current dataset.

For classical SDM transmission, the demonstrated precision far exceeds
digital signal processing buffer requirements, enabling per-span ICS monitoring compatible
with deployed telecom infrastructure at $1550\,$nm. For quantum
networks, the ability to characterize ICS with sub-picosecond
resolution enables compensation of timing mismatch before it degrades photon
interference in MCF-based links, and the
complementary pulsed-laser regime extends this capability to
field-deployed fibers with high channel loss. The Fisher information
framework established here provides a systematic basis for optimizing
photon bandwidth, scan parameters, and delay calibration to approach
the CRB in future experiments.

\appendix
\section{Statistical Inference for the ICS Scaling Exponent}
\label{app:inference}

The RMS ICS data are fitted to the general power-law model
\begin{equation}
  \sigma_\tau(L) = A\,L^{\alpha},
  \label{eq:general_scaling}
\end{equation}
introduced in the results section. The exponent $\alpha$ and its
uncertainty are estimated by weighted least squares in log space,
with weights $w_i = 1/\sigma_{y,i}^2$ where
$\sigma_{y,i} = \sigma_{{\rm ICS},i}/({\rm ICS}_i\cdot\ln 10)$
are the propagated log-space uncertainties; length points with fewer
accessible core-pair combinations are additionally down-weighted.
Robustness is assessed by non-parametric bootstrap resampling
($N = 5000$ datasets with replacement); the 95\% confidence interval
is obtained by the percentile method. Full details follow standard
practice~\cite{kay1993fundamentals}. 

\begin{backmatter}

\bmsection{Funding}
This research was funded by Fondo Nacional de Desarrollo Cient\'{\i}fico y
Tecnol\'ogico (FONDECYT) Regular (Grant Nos.\ 1240746, 1231940, 1260111, 1240843), ANID -- Millennium Science
Initiative Program -- ICN17\textendash012, and ANID Anillo Project ATE250003. LLT was also suported by project UCO 1866. LMF was supported by Conselho Nacional de Desenvolvimento Científico e Tecnológico (CNPq - DOI 501100003593). MN was supported by ANID BECAS/Magister Nacional 2021-22211554, Government of Spain (Severo Ochoa CEX2019-000910-S, FUNQIP and NextGeneration EU PRTR-C17.I1) and European Union’s Horizon Europe research and innovation programme under the MSCA Grant Agreement No. 101081441. SG was supported by  FONDECYT Grant No. 3210359. 

\bmsection{Acknowledgments}
The authors thank P.H. Souto Ribeiro for useful conversations.

\bmsection{Disclosures}
The authors declare no conflicts of interest.

\bmsection{Data availability}
Data underlying the results presented in this paper are not publicly
available at this time but may be obtained from the authors upon
reasonable request.

\end{backmatter}


\end{document}